\documentclass[10pt,conference]{IEEEtran}

\usepackage[ruled]{algorithm2e}
\usepackage{listings}
\usepackage{multirow}
\usepackage{cite}
\usepackage{color}
\usepackage{array}
\usepackage{url}

\usepackage{tikz}
\usepackage{scalefnt}
\usetikzlibrary{arrows,automata,shapes,snakes,petri}
\usetikzlibrary{positioning}
\usepackage{pifont}

\usepackage{epstopdf}
\usepackage{subcaption}
\usepackage{booktabs}
\usepackage[normalem]{ulem}
\definecolor{javagreen}{rgb}{0.25,0.5,0.35} %

\SetAlFnt{\small}
\SetAlCapFnt{\small}
\SetAlCapNameFnt{\small}
\SetAlCapHSkip{0pt}
\LinesNumbered
\IncMargin{-\parindent}
\selectfont

\begin{document}

	\title{Generating Predicate Callback Summaries \\ for the Android Framework}

     \author{\IEEEauthorblockN{Danilo Dominguez Perez and Wei Le}
        \IEEEauthorblockA{Iowa State Univeristy\\
            \{danilo0,weile\}@iastate.edu}
    }
	
	\date{}
	
	\maketitle
	\begin{abstract}
  	One of the challenges  of analyzing, testing and debugging Android apps is 
  	that the potential execution orders of callbacks are missing from the apps' 
  	source code. However, bugs, vulnerabilities and refactoring 
  	transformations have been found to be related to callback sequences. 
  	Existing work on control flow analysis of Android apps 
    have mainly focused on analyzing GUI events. GUI events, although being a key part of 
    determining control flow of Android apps, do not offer a complete 
    picture. Our observation is that orthogonal to GUI events, the Android API 
    calls also play an important role in determining the order of callbacks. 
    In the past, such control flow information has been modeled manually. 
    This paper presents a complementary solution of constructing program paths 
    for Android apps.  We proposed a specification technique, called {\it Predicate 
    Callback Summary} (PCS), that represents the callback control flow 
    information (including callback sequences as well as the conditions under 
    which the callbacks are invoked) in Android API methods and developed 
    static analysis techniques to automatically compute and apply such 
    summaries to construct apps' callback sequences. Our experiments show that by applying PCSs, we are able to 
    construct Android apps' control flow graphs, including inter-callback 
    relations, and also to detect infeasible paths involving multiple callbacks. Such 
    control flow information can help program analysis and testing tools to report 
    more precise results.
    Our detailed experimental data is available at: 
  	\url{goo.gl/NBPrKs}

\end{abstract}

	
	


\section{Introduction}
Android apps are implemented by overriding callback methods from building block classes in the Android framework such as \textit{android.app.Activity} and  \textit{android.app.Service}. One challenge for analyzing, testing and debugging such software is that the program paths, the sequence of callbacks potentially 
executed at runtime, are missing from apps' source code. On the other hand, program paths of callback sequences are essential for any inter-procedural analysis implementation for Android apps. For example, bugs and vulnerabilities have been found along paths involving multiple callbacks~\cite{arzt2014flowdroid} 
\cite{pathak2012keeping} \cite{Hsiao:2014:RDE:2594291.2594330}. Without path information, we will have difficulties to detect or diagnose such bugs. Control flow information of callbacks is also useful for 
verifying code transformations such as automatic refactoring techniques. There has been research 
which shows that refactoring of Android asynchronous constructs involve 
multiple related callbacks~\cite{dannydig}. 

In the past, control flow analysis of Android apps have mainly focused on 
analyzing GUI behaviors and determining the order of callbacks related to GUI events. 
Our observation is that orthogonal to GUI events, the Android API methods also play an important role in determining the order of callbacks implemented in the apps. We analyzed 1000 apps from F-Droid~\cite{fdroid} and the Google play market, and found that an app has, on average, 1204 API calls and up to 4029 API calls. Additionally, among the top 100 frequently invoked API calls, 47\% of them contain callbacks and 43\% contain multiple callbacks. Without 
considering API methods, we are unable to obtain all possible paths of 
callbacks for the apps. Due to its importance, previous studies have used manual approaches to model the control flow of callbacks in the Android API methods~\cite{yang-icse15} \cite{yang2015static} \cite{wang2016unsoundness}. The challenges are that the Android framework evolves fast, and it contains more than 20,000 API methods. Manual approaches are infeasible and unable to keep up with the rapid evolution of the Android framework.  

This paper presents a complementary solution to GUI models for identifying 
control flow of callbacks implemented in apps. We proposed a specification technique, 
called {\it predicate callback summary (PCS)}, and static analysis 
techniques that automatically compute the PCSs from the Android API methods. 
A PCS is a control flow--based graph representation that extracts from the API 
implementation 1) all the callback sequences, 2) the predicates depending on 
which the callbacks are executed, and 3) the updates that can determine the 
outcome of the predicates. The callback sequences in PCSs enable inter-procedural analysis for Android apps and help developers understand how their code (callback methods) is executed in the framework. The predicate and update nodes in the PCSs are used to determine the feasibility of program 
paths, which are very important for eliminating false positives in static 
analysis and for improving the evaluation of test coverage. The predicate 
nodes are also useful for generating test inputs that can exercise interesting 
callback sequences, e.g., testing the interface between the app and the Android 
framework.

The challenges to compute PCSs are twofold. First, we need to identify how 
synchronous and asynchronous callbacks are invoked in the Android framework and extract their order from the existing control flow of the API implementation. 
Second, to ensure that PCSs can be directly applied in the app without 
repeatedly analyzing the API method at each API call site, we need to abstract 
away all the local variables to a representation that is visible to 
API clients. The two tasks require both control flow and symbolic analysis 
of the Android framework. The framework contains about 8 millions lines of 
code, and the call graph for each API method can be considerable large, containing call chains with hundreds of calls. Is extremely hard to achieve a balance between scalability and precision of static and symbolic analysis for such code base. 

To demonstrate that PCSs are useful, we developed a client analysis that can automatically plug in PCSs at API call sites and construct {\it inter-callback inter-procedural control flow graphs (inter-callback ICFGs)} for Android apps. We also adapted an existing infeasible path detection algorithm~\cite{bodik1997refining} for the apps' control flow graphs and PCSs to show the usefulness of predicate and update nodes in PCSs. 

These static analysis algorithms were implemented in a tool called \textit{Lithium}. We generated PCSs for the top 500 frequently used API methods on the Android Framework~5.1. Our experiments show that generating a PCS takes 34.5 seconds on average. Compared to the ICFGs of the API methods, the summaries are 99\% smaller on average.  Using the PCSs, we are able to connect sequences of up to 44 callbacks in the app and also able to identify the infeasible paths that contain multiple callbacks. We also verify the feasibility of the paths of callback sequences generated by comparing them against dynamically generated traces. 

To the best of our knowledge, this is the first work on summarizing control 
flow of callbacks in API methods. Although our implementation and 
studies are mainly based on the Android framework, the techniques may be 
generally applicable for summarizing any library that heavily invoke 
callbacks. PCSs are a complementary solution to existing GUI based control flow 
analyses. By combining both models, we can obtain a complete inter-callback 
ICFG for the app, enabling any path-based static and dynamic analysis. The 
contributions of the paper can be summarized as follows: 

\begin{itemize}
\item a novel specification, PCS, for representing control flow of callbacks in an API method
\item static analysis techniques and a tool, {\it Lithium}, to compute and apply such summaries to construct apps' inter-callback ICFGs and exclude infeasible paths, and 
\item the engineering efforts and the experience of analyzing large real-world libraries; the experimental results that demonstrate the scalability, precision and usefulness of the techniques.
\end{itemize}


The rest of the paper is organized as follows. In Section \ref{sec:motivation}, we present the motivation of the work. In Sections~\ref{sec:define} to~\ref{sec:apply}, we show how we define, compute and apply
PCSs. We present our implementation and experimental
results in Section~\ref{sec:results} and a discussion of limitations in Section~\ref{sec:discuss}, followed by related work in Section~\ref{sec:related}, and the conclusions in Section~\ref{sec:conclusions}.


\section{A Motivating Example}~\label{sec:motivation}
\subsection{The Challenge}
\begin{figure}[ht]
	\resizebox{0.42\textwidth}{!}{
		\lstset{basicstyle=\footnotesize, numbers=left, numberstyle=\tiny, language=Java}
\begin{lstlisting}
 class HostListActivity extends Activity {
  public void onStart() {
    this.startService(new Intent(this, 
      TrackingRecordingService.class));
    this.bindService(new Intent(this, 
      TrackingRecordingService.class)),...);
  }
  public void onStop() {
    super.onStop();
    this.unbindService(connection);
  }  
}  
class TrackingRecordingService extends Service {
  public void onCreate() { 
    wifilock.acquire(); 
  }
  public boolean onUnbind(Intent intent) {
+   if (bridges.size() == 0) this.stopSelf();//patch
+   return true;
 }
  public void onDestroy() {
    if (wifilock != null && wifilock.isHeld())
       wifilock.release();
  }
}
\end{lstlisting}
	}
	\caption{ConnectBot Bug\label{fig:connectbot93}}
\end{figure}

In Figure~\ref{fig:connectbot93}, we show a code snippet adapted from the app {\tt ConnectBot}. This snippet contains a {\it no-sleep} bug~\cite{pathak2012keeping} that can drain the phone battery and spans through multiple callbacks. At lines~3--6, the app invokes the Android API methods {\tt startService} and {\tt bindService} in {\tt HostListActivity} to start and bind the service {\tt TrackRecordingService}. As a result of the API calls, {\tt onCreate} at line~14 is invoked, and the {\it WiFi lock} is acquired (see line~15). When the app needs to unbind the service, it invokes {\tt unbindService} at line~10. This Android API method calls {\tt onUnbind} if the service is started, or {\tt onUnbind} followed by {\tt onDestroy} if the service is not started. 

The bug occurs because the developer assumed that {\tt onDestroy} is always invoked after {\tt unbindService} is called, independent of whether the service is started. The consequence is that the service is never destroyed and the {\it WiFi lock} acquired at line~15 is never released. The bug is fixed by calling {\tt stopService} or {\tt stopSelf} to stop the execution of the service (see line~18).

In this example, we show that to detect such bug, we need to know the control flow between multiple callbacks. The challenge is that developers have no direct control on the order of the callbacks in their apps' source code. For example, it is the Android API methods {\tt startService} and {\tt bindService} invoked in {\tt onStart} (see lines~3 and 5) that determine the order of the callbacks {\tt onCreate}, {\tt onUnbind} and {\tt onDestroy} implemented in {\tt TrackingRecordingService}. Considering the large number of API methods and the complexity of each API method implemented in the framework, it is challenging for developers to remember which callbacks are invoked in which API methods, including the order of callbacks and the conditions that need to be true for their execution. Without such knowledge, it is easy to make mistakes and introduce bugs. We need a plug-in-and-go type of model for API calls to help developers understand the control flow of calls in the Android API methods and improve the scalability and precision of static analysis and testing tools.


\subsection{Our Goal}
We aim to define and automatically infer PCSs to summarize the control flow of callbacks implemented in libraries such as the Android framework. Using API calls {\tt startService} and {\tt unbindService} shown in Figure~\ref{fig:connectbot93} as examples, we explain what a PCS is and how it can help to construct apps' control flow graphs that interconnect callbacks.

\begin{figure}[!ht]
	
	\centering
	\includegraphics[width=0.95\columnwidth]{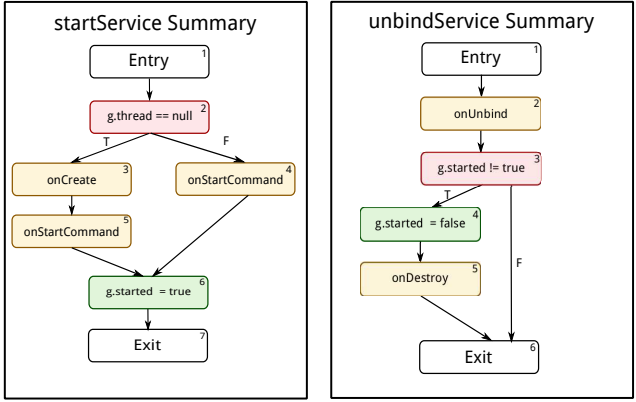}
	\caption{PCSs for {\tt startService} and {\tt unbindService}~\label{fig:example}}
\end{figure}

Figure~\ref{fig:example} shows the PCSs for {\tt startService} and {\tt bindService} (simplified versions of the original graphs). The figure shows that 1) by traversing the PCSs, we can obtain sequences of callbacks; 2) callbacks are guarded by predicates; 3) the PCSs include nodes (green nodes) that may modify values used in predicates, and such updates can have a global effect on other API methods by changing the outcome of conditions that determine the invocation of callbacks.

Specially, the PCS for {\tt startService} shows that there are two potential paths that execute the callbacks: when the predicate {\tt g.thread == null} is true, path $\left<2, 3, 5, 6\right>$ is executed and {\tt onCreate} is invoked asynchronously, followed by {\tt onStartCommand}; otherwise, only the asynchronous call {\tt onStartCommand} is invoked along path $\left<2, 4, 6\right>$. By analyzing the PCSs for both API methods, we found that the update  {\tt g.started = true} in {\tt startService} at node~6 has an impact on the invocation of the callbacks in {\tt unbindService}. Shown on the right in Figure~\ref{fig:example}, the PCS contains a predicate node (node 3) that tests if {\tt g.started != true}. The outcome determines if the callback {\tt onDestroy} will be executed. Using the two PCSs, we can detect the mistake in Figure~\ref{fig:connectbot93}, knowing that as long as {\tt startService} is invoked, {\tt g.started} is set to true. As a result, node~3 in {\tt unbindService} always returns false, and {\tt onDestroy} is not invoked. 

The PCSs can be used to automatically construct apps' control flow graphs that interconnects multiple callbacks; we call them {\it inter-callback ICFGs}.
\begin{figure}[!ht]
	
	\centering
	\includegraphics[width=\columnwidth]{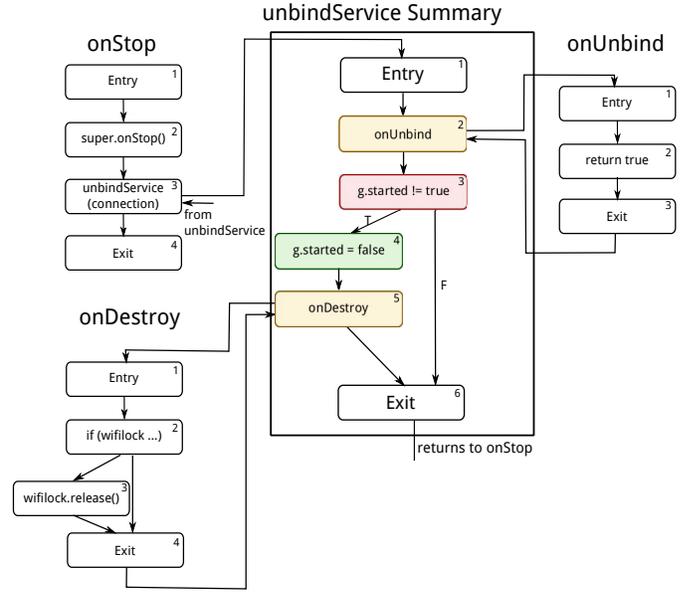}
	\caption{inter-callback ICFG related to {\tt unbindService}~\label{fig:appicfg}} 
\end{figure}
In Figure \ref{fig:appicfg}, we show an example of an inter-callback ICFG for the code snippet in Figure~\ref{fig:connectbot93}. Due to space limitation, we only provided the inter-callback-ICFG related to {\tt unbindService} call. It shows the ICFG for the three callbacks {\tt onStop}, {\tt onDestroy} and {\tt onUnbind}. Before the application of the PCS for {\tt unbindService}, the three ICFGs are disconnected and we cannot obtain the paths that traverse all the three callbacks. At node~3 in {\tt onStop}, the API method {\tt unbindService} is invoked. We plug in its PCS by constructing an edge from the call site at node~3 in {\tt onStop} to the entry of {\tt unbindService} and another edge from the exit of PCS back to the call site. In the PCS, we discover that the callbacks {\tt onUnbind} and {\tt onDestroy} are invoked. We find the corresponding callbacks in the app and add edges to connect to the two callbacks, as shown in Figure~\ref{fig:appicfg}. 
Using a similar approach, we can also construct an inter-callback ICFG to connect {\tt onStart} and {\tt onCreate}. Combining our solution with GUI-based control flow graph construction~\cite{yang-icse15, yang2015static}, we can further sequence {\tt onStart} and {\tt onStop} and obtain a complete faulty path for the app. We leave this for future work. 



\section{Defining PCS}~\label{sec:define}
A Predicate Callback Summary (PCS) is an interprocedural abstraction for library methods (in our case for methods of the Android API) that keep the control flow information about the callbacks that are invoked and the context needed for their invocation. More formally, a PCS is a directed graph 
$G = (N_c \cup N_p \cup N_u, E)$ where $N_c$ is the set of \textit{callback nodes}, $N_p$ is the set of \textit{predicate nodes} and $N_u$ is the set of statements which update values of variables used in \textit{predicate nodes}; called \textit{update nodes}. $E$ is the set of edges between any two nodes in the set $N_c \cup N_p \cup N_u$.
PCSs aim to summarize all potential execution paths of the callbacks implemented in library methods. In the rest of the section, we explain each type of node.

\vspace{0.2cm}
\noindent{\bf Callback Nodes.} 
The framework executes callback methods through objects passed from the app that are instances of classes extending the Android API. There are two ways an app can pass the object receiver of a callback to the framework. Commonly, the object receiver can be passed through the parameters (including fields of the parameters) of an API call. The object receiver can also be the calling object of an API call. For example, when the object receiver is an instance of a class extended by the app and the API call executes a method that was overridden by the app. A callback node $n \in N_c$ represents a callback call site that is executed in an API method. The two key pieces of information specified in callback nodes are 1) the object receiver of the callback, and 2) the callback signature. This information is useful to identify the correct callback methods in the app during client analysis.


\vspace{0.2cm}
\noindent{\bf Predicate Nodes.} A predicate node $n \in N_p$ provides the conditions that need to be satisfied for the execution of a callback or a sequence of callbacks. 
They are useful, along with update nodes, to determine the feasibility of a sequence of callbacks to be executed at an API call site. 

The predicate nodes are abstracted to boolean expressions of type $a \; op \; b$ where $op \in \{<, >, \leq, \geq, ==, !=\}$ and $a$ and $b$ can be constants, {\it abstract variables} or arithmetic expressions on constants and abstract variables. For instance, in Figure~\ref{fig:example}, {\tt g.thread == null} at node~2 in {\tt startService} and {\tt g.started != true} at node~3 in {\tt unbindService} are predicate nodes. In these two boolean expressions, the left operand is abstracted to abstract variables ({\it static, g, thread}) and  ({\it static, g, started}), and the right operands are resolved to constants (null and boolean constant respectively). 

\vspace{0.2cm}
\noindent{\bf Update Nodes.} Once we have all the predicate nodes, we need to find all the statements that can affect the outcomes of these predicates. Statically, an update node $n \in N_u$ is an assignment whose left side is a variable used in one of the predicate nodes found. These assignments can contribute completely or partially to change the outcomes of predicate nodes. Dynamically, the update nodes are the ones that can change the program state and impact the invocations of succeeding callbacks located in the current API method or the succeeding API methods. 

	
 
For example, the Android framework keeps a map of all the \texttt{Service} objects running in an app. The map is created when the app starts, and it can be modified and accessed throughout the app's lifetime using a static variable defined on the framework. When any \texttt{Service} object is going to be started, the app invokes {\tt startService}. This API call first checks if the {\tt service} object is already started by inquiring the static variable. If the object does not exist, the app creates the corresponding service by calling the {\tt onCreate} callback and also updates the state of the service stored in the static variable. In this case, predicate conditions on the static variables are predicate nodes and the statements that change its values are update nodes. 

 \begin{figure}[h]
 	\centering
 	\resizebox{0.45\textwidth}{!}{
 		\lstset{basicstyle=\footnotesize, numbers=left, numberstyle=\tiny, language=Java}
\begin{lstlisting}[lastline=26]
class LoaderManager {
  Loader<D> initLoader(int id, Bundle args, LoaderCallbacks<D> c) {
    LoaderManager r0 = this;
    boolean creatingLoader = r0.mCreatingLoader
    if (creatingLoader == true) {
      throw new IllegalStateException("...");
    }    
    LoaderInfo info = r0.mLoaders.get(id);
    if (info == null) {
      info = r0.oldLoader;
      r0.mLoaders.put(id, info);
      c.onCreateLoader(); // callback call
    }
    boolean haveData = info.mHaveData;
    reset(haveData); 
  }
  void reset(boolean haveData, LoaderCallbacks<D> c) {
    if (haveData == true) {
      c.onReset();
    }
  }
}
\end{lstlisting}
 	}
 	\caption{\texttt{LoaderManager.initLoader}\label{fig:preconditions}}
 \end{figure}

One of the requirements for the PCSs is that they need to be directly plugged in at any API call site and analyzed with apps' code without any re-computation of the API method. This requires that the variables and expressions used in the three types of nodes are also visible to the client or to the other API methods invoked by the client. For that, we need an abstraction for the local variables used in predicates. In our analysis, each local variable is resolved to a set of aliases represented by access paths~\cite{sridharan2013alias}. The access paths are resolved in a backward analysis from the predicate node until a visible variable (they are either the calling object of the API method, a parameter of the API method or a static variable ) is obtained.
  
As an example, see the code snippet of the API method {\tt initLoader} in  Figure~\ref{fig:preconditions}. At line~19, a callback {\tt onReset} is invoked through an object receiver {\tt c}, and the predicate is {\tt haveData == true} (see line~18). However,  {\tt haveData} is a local variable and is not visible outside the API method {\tt initLoader}. Presenting such information in the PCS does not help to determine whether {\tt haveData} is true and whether the callback {\tt onReset()} will be invoked at the API call site. Therefore, in the PCS, we map the local variable {\tt haveData} to an abstract variable.
For instance, along the path that traverses lines~4, 10 and 14, we obtain an abstract variable represented using a three tuple ({\it calling object, LoaderManager, oldLoader.mHaveData}). The tuple indicates that the variable is computed from the calling object of the API method, with the class type~\texttt{LoaderManager}, and we can compute it by first accessing its field {\tt oldLoader} and then {\tt mHaveData}.

Generally, in the three tuple that specifies the abstract variable, the first element specifies the visibility and scope of the variable and has the domain of {\it static variable} (from the classes of the Android framework), the API {\it calling object} or the {\it input parameters} of the API method.  The second element of an abstract variable specifies the class type of the variable. The third element provides the details on how to compute the values for the variable, e.g., via an access path.



	\section{Computing PCS}\label{sec:compute}
\begin{figure}[!ht]
	\centering
	\includegraphics[width=\columnwidth]{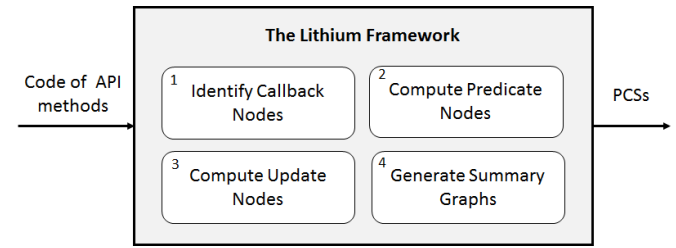}
	\caption{An Overview of Lithium}\label{fig:framework}
\end{figure}

In this section, we present our algorithms to compute PCSs from the source code of Android API methods. Figure \ref{fig:framework} shows our static analysis framework, Lithium, which have 4 main phases: identify callback nodes, compute predicate nodes, compute update nodes, and generate the summary graphs. The framework uses ICFGs generated from the source code of the API methods as input. 

\subsection{Identifying Callback Nodes}\label{sec:identifycallbacks}
The goal of this phase is to locate callback call sites in API methods. Specifically, we aim to identify the program paths that can reach the callbacks from the entry of the API method. Along these paths, we then can generate summaries for predicate nodes and update nodes. 


\subsubsection{Searching for Callback Call Sites}

In the first step, we address the question of {\it which calls invoked in an API method can be callbacks?} We identify a list of potential callback signatures by analyzing the class hierarchy and the visibility of the methods in the Android framework. Specifically, we find every non-static, non-final method of a non-final class that has a visibility of public or protected. Additionally, we also identify all the methods of public interfaces in the Android framework. These methods can be overridden by the apps and executed in the Android framework via dynamic dispatch and thus they can be callbacks.

Once we have the list of potential callback signatures, we inspect all the call sites in the framework and verify if they match one of the callback signatures. If a callback signature is matched, we perform a backward traversal from the call site on the call graph until an API method is reached. This step can generate a set of \textit{call chains} of the form $\{m_0, ..., m_n\}$ where $m_0$ is an API method followed by a sequence of framework method calls until the callback $m_n$ is reached. We define $\mathcal{C}$ to be the set of all the call sites that match a callback signature and the function $Chains(c)$, for given a call site $c \in C$, return the set of all call chains found that reach $c$.
We use these call chains in the next phases of our analysis to ensure that it only traverses the paths of interest for scalability.


\paragraph{Handling Asynchronous Callbacks} One of the challenges to identify the call chains is that there are callbacks that are invoked implicitly through a message passing mechanism developed in Android\cite{handlerthread}. The message passing mechanism is implemented using the \texttt{Handler} class on the framework through a pair of methods \texttt{sendMessage} and {\tt handleMessage}. At runtime, when {\tt sendMessage} is invoked from a {\tt Handler} object in the API method, a message is posted into an event queue.  When the message is dispatched, {\tt handleMessage} from the same {\tt Handler} class is invoked to handle the message.  The {\tt handleMessage} method can invoke a set of callbacks.



To include such asynchronous callbacks into the summary, we need to construct an edge to connect  \texttt{sendMessage} and its corresponding {\tt handleMessage} in the ICFG. To do so, we first determine the class name (type) of the object receiver for every asynchronous call of \texttt{sendMessage}. We then find the {\tt handleMessage} method defined in the class. For example, in Figure~\ref{fig:asynchronous}, according to line~10,  the object receiver of {\tt sendMessage} at line~12 has a type {\tt ActivityThreadHandler}. Thus, we identify that {\tt handleMessage}, implemented in the class {\tt ActivityThreadHandler} at line~2, is a match and we create an implicit edge from \texttt{sendMessage} to \texttt{handleMessage} in the framework's call graph. Any callback invoked in this method should be linked to the asynchronous call site of {\tt sendMessage} at line~12.



\begin{figure}[h]
    \centering
    \resizebox{0.45\textwidth}{!}{
        \lstset{basicstyle=\footnotesize, numbers=left, numberstyle=\tiny, language=Java}
\begin{lstlisting}[lastline=26]
class ActivityThreadHandler extends Handler {
  public void handleMessage(Message m) {
    switch (m.what) {
    case LAUNCH_ACTIVITY:
      handleStartActivity(...); break;
    }
  }
}
class Activity {
  ActivityThreadHandler threadHandler;
  public void startActivity(...) {
    threadHandler.sendMessage(LAUNCH_ACTIVITY);  
  }
}
\end{lstlisting}
    }
    \caption{handleMessage in ActivityThreadHandler}\label{fig:asynchronous}
\end{figure}

%
\subsubsection{Resolving Object Receivers}\label{sec:objectreceivers}
In this step, we identify the object receiver of the call sites found in the previous step. Algorithm~\ref{alg:resolvingreceivers} shows the steps to resolve the object receivers for each call chain and returns the set of callback nodes $N_c$. The procedure takes as input the set of all the call sites that match a callback signature $\mathcal{C}$ and the $ICFG$ generated for the framework. At line~3 we find aliases for the base object of the call sites using a demand-driven alias analysis used in \cite{arzt2014flowdroid}. At line~6 we verify if the base object of the call site (first element in call chain) is alias of the caller object in the API method. At line~7 we add the \textit{this} (caller) reference to the receivers. Then, we verify the parameters of the API call. The function \texttt{Params} returns the parameters references of the API call. We evaluate each parameter and its fields (in case they may point to the object receiver of the callback) using the function \texttt{MatchParam}. We use \texttt{Unknown} for the case when the object receiver is not resolved to the caller or a parameter. This can happen when the object receiver is passed through a different API method and is stored in the framework. For example, the method \texttt{TextView.setText} executes callbacks for objects of type \texttt{TextWatcher} which are stored in a list of listeners updated by the API method \texttt{TextView.addTextChangedListener}. There is also the case when the object receiver is internal to the framework and it cannot point an object passed from the app, which means it cannot be a callback method. We leave the detection of such false positives for future work.

\begin{algorithm}
    \caption{Resolving Object Receivers}
    \label{alg:resolvingreceivers}
    \small
    {\scalefont{0.95}
        
        \SetKwFunction{backwardAliasAnalysis}{BackAliasAnalysis}
        \SetKwFunction{chains}{Chains}
        \SetKwFunction{head}{Head}
        \SetKwFunction{base}{Base}
        \SetKwFunction{isEmpty}{IsEmpty}
        \SetKwFunction{matchParam}{MatchParam}
        \SetKwFunction{params}{Params}
        \SetKwFunction{isInternal}{IsInternal}
        \SetKwInOut{Input}{Input}
        \SetKwInOut{Output}{Output}
        \SetKwComment{Comment}{start}{end}

        \Input{$C, ICFG$} 
        \Output{$O: $ set of of callback nodes with their call chains}
        
        \ForEach{$c \in C$}{
            \ForEach{$chain \in $ \chains{$c$}}{
                $S = \backwardAliasAnalysis(\base{c}, chain, ICFG)$\;
                $receivers = \{\}$\;
                \ForEach{$p \in S$}{
                    \If{$p$ may point to $\base{\head{chain}}$}{
                        $receivers = receivers \cup \{this\}$
                    }
                    \ForEach{$param \in \params{\head{chain}}$}{
                        \If{$p$ may point to $param$ or it fields}{
                            $receivers = receivers \cup \matchParam{param, s}$
                        }
                    }
                }
                \lIf{$\isEmpty{receivers}$}{ 
                    $receivers = \{unknown\}$
                }
                $O = O \cup ((c, receivers), chain)$
            }
        }          
        \Return{$O$}
    }
\end{algorithm}



\subsection{Computing Predicate Nodes}\label{sec:identifypreconditions}
Our approach here is to first perform control flow analysis to identify conditional branches that a callback node is transitively control dependent on, and report them as \textit{predicate nodes} (Section \ref{sec:findingpreconditions}). We then resolve the local variables contained in the predicate nodes to symbolic expressions of abstract variables via a backward symbolic substitution (Section \ref{sec:resolvingvisiblevars}). 

\subsubsection{Identifying Predicate Nodes}\label{sec:findingpreconditions}

Predicate nodes are the conditional branch statements in the ICFG of the API method that decide whether a callback should be executed. For each method $m_i$ appeared in the call chain $\{m_0, ..., m_n\}$ (computed in the phase {\it Identifying Callback Nodes}), we applied a control flow analysis shown in Algorithm~\ref{alg:controlnodes}.

The inputs of the algorithm are the CFG of the method $m_i$,
and the program point of interest $p$. Depending on $m_i$, $p$
can either be the callback $c$ or the call site of the method $m_{i+1}$
in the call chain. The algorithm reports a set of branch nodes
which $p$ is transitively control dependent on.
At line 2, the algorithm traverses every conditional branch
in the method. At line 3, \texttt{Influence(b)} returns all the statements in the CFG that are transitively control dependent on
the conditional branch statement $b$ \cite{weiser1981program} \cite{Weiss:1992:TCC:151333.151337}. If the statement of interest $p$ is one of such statements, the branch node $b$ will
be stored in the results $B$.
%

%
%
%
%
\begin{algorithm}
    \caption{Identifying Predicate Nodes}
    \label{alg:controlnodes}
    \small
    {\scalefont{0.85}
        
        \SetKwFunction{infl}{Influence}
        \SetKwInOut{Input}{Input}
        \SetKwInOut{Output}{Output}

        \Input{$m_i: CFG = \langle N, E \rangle$, $p \in N: $ callback $c$ or call site of $m_{i+1}$} 
        \Output{$B \subset N: $ the set of conditional branches considered predicate nodes}
        
        $B = \{\}$ \\
        \ForEach{conditional branch $b \in m_i$}{
            $S = $ \infl{$b$} \\
            \lIf{$p \in S$}{
                $B = B \cup {b}$
            }
        }  
        
        \Return{$B$}
    }
\end{algorithm}

\subsubsection{Summarizing Predicate Nodes}\label{sec:resolvingvisiblevars}
The goal of this step is to convert any of the local variables in the predicate nodes to be \textit{abstract expressions}. The approach we used is demand-driven, symbolic back-substitution \cite{bodik1998path}. We unroll loops once to assure termination of the back-substitution algorithm. The analysis starts at each predicate node and propagate backwards along all paths reachable from the predicate nodes. At any assignment that defines the local variables under tracking, we update the variables symbolically. For local variables, this substitution can generate a set of access paths. The analysis ends when the variable is resolved to an expression of abstract variables or constants, or when the traversal reaches the entry of the API method. We use these abstract variables to identify update nodes (see Section~\ref{sec:identifypostconditions}).


As an example, in Figure \ref{fig:preconditions}, the local variable {\tt haveData} in the predicate at line~18 can be resolved to \{({\it calling object, LoaderManager, oldLoader.mHaveData}) $\vee$ ({\it calling object, LoaderManager, mLoaders.get().mHaveData})\}, where $\vee$ indicates the merge of the abstract variables along the path at lines~14, 10 and 4, and along the path at lines~14,~8~and~4.

From the above example, we show that at each statement determined to be relevant to the variable under tracking, we update the symbolic expressions. In case of a field dereference, we add the field to the access path under tracking and keep inquiring whether the base object can be resolved to static variables, input parameters or calling objects. Due to the unprohibited expenses, our current implementation does not traverse to callee methods when it reaches a method call and treat them as if they were fields, adding them to the computed access path together with field references (see the above example {\tt mLoaders.get().mHaveData}). Our insight is that the expressions on abstract variables in the predicate nodes are used together with the update nodes to decide infeasible paths, and in our empirical studies, we found that among one the most common cases where local variables were computed to return values, the methods belonged to collection classes (\textit{JDK} and Android collections); therefore, we manually define the templates for these classes given their side effects, and use them to resolve the correlations between predicate and update nodes for infeasible paths.


In Table \ref{tab:templates}, we show a few example of the templates we developed. Under {\it Class}, we show to which class the methods belong to. Under {\it Predicate Nodes}, we list the method calls frequently used in predicate nodes. The column provides a set of methods used to test the state of a collection. For example, {\tt isEmpty} tests if a collection is empty. Under {\it Update Nodes}, we list the types of statements that potentially can answer questions regarding the conditions in the predicate node---these are the methods that have side effects on the collections. Note that the methods whose names end with a star (*)  represent a set of similar methods whose names start with the same prefix. For example, we have the method {\tt contains} to test if a member belongs to a collection, and {\tt containsAll} tests if a set of values belong to a collection. We represent such calls using {\tt contain*}.

 \begin{table}
 	\centering
 	\caption{Matching Predicate Nodes and Update Nodes}\label{tab:templates}
 	\resizebox{0.49\textwidth}{!}{
 		\begin{tabular}{|l|l|l|} \hline
 			Class         & Predicate Nodes & Update Nodes  \\\hline \hline
 			\multirow{2}{*}{java.util.List} & isEmpty, size, & \multirow{2}{*}{add*, remove*, set} \\
 			& get*, contains* & \\ \hline
 			java.util.Set & isEmpty, size, contains* & add*, remove* \\ \hline
 			\multirow{2}{*}{java.util.Map} & isEmpty, size, &  \multirow{2}{*}{put*, remove} \\
 			& contains*, get & \\ \hline
 			\multirow{2}{*}{android.util.ArrayMap} & isEmpty, size & setValueAt, put*, \\
 			& value*, contains* & remove* \\ \hline
 			\multirow{2}{*}{android.util.SparseArray} & \multirow{2}{*}{size, value*} & setValueAt, put*,  \\
 			& & remove*, delete \\ \hline
 		\end{tabular}
 	}
 \end{table}
 
 

\subsection{Computing Update Nodes}\label{sec:identifypostconditions}

The goal of including update nodes in the PCSs is to determine under a particular condition, which callbacks (either in the current API method or in the succeeding API methods) should be invoked. 
To determine which statements in the API method can be update nodes, we need to 1) obtain all the abstract variables used in the predicate nodes for all the Android API methods, and 2) find the assignments and the framework calls that can potentially change the values of these variables.

We start by analyzing assignment statements in the API methods and summarizing the destination of the assignments into abstract variables and their expressions. We terminate the analysis for this assignment early if we found that the access path we obtained so far does not match any of the abstract variables in the list. In the second step, we examine the method calls to identify possible matches with templates defined in Table \ref{tab:templates}. When a method call signature matches a template, we determine if the calling object of this call is resolved to any abstract variables of interest. 

Following the example in Figure \ref{fig:preconditions}, the abstract variable ({\it calling object, LoaderManager, mLoaders.get()}) used in defining the predicate can be paired with the method call at line~11, as the latter can be resolved to ({\it calling object, LoaderManager, mLoaders.put()}).  Let us suppose {\tt initLoader} has been invoked twice along the path: in the first API call, at line~11, we invoke {\tt put} on the calling object's field {\tt mLoaders}, and we then know in the second invocation of {\tt initLoader}, at line~8, the call {\tt get} on the same calling object will return a non-null value, and the branch at lines~9--13 will not be executed. The paired methods {\tt get} and {\tt put} helps resolving the branch correlation in the two API calls.

 


\subsection{Generating Summary Graphs}\label{sec:gensummary}

In the first three phases, we mark all the identified callback nodes, predicate nodes and update nodes on the ICFG of the API method. To add edges between the marked nodes and generate the summary graph, we traverse the ICFG and determine the reachability between the marked nodes in the ICFG with the goal that the final summary graph should keep the original control flow between the marked nodes.

Algorithm \ref{alg:summarygraph} takes an input the ICFG of the API method with the three types of marked nodes and generates a summary graph for the API method. The $worklist$ at line~2 stores a pair of nodes, $q$ and $n$, where $q$ is the last marked node seen and $n$ is the node encountered during the traversal of the ICFG. This pair of nodes are always reachable from each other on the ICFG, as the two nodes initially are the same node (see line~3). When creating a new worklist element at line~12, the successors of $n$ are replaced, which are still able to reach $q$. At lines~7 and 8, when we find that $n$ is a marked node, an edge is added between $n$ and $q$.

A key function {\tt Succ} at line~12 handles the challenges of interprocedural analysis. If $n$ is the exit of the current method, the successors are located at the next statement of its call sites. If $n$ is the call site, the successors can be found at the entry of its callees (only the ones containing marked nodes). A special case when a query reaches {\tt sendMessage} in the ICFG, we find its successors in the inlined callbacks at the call site of {\tt sendMessage} (see Section 4.1).

\begin{algorithm}
	\caption{Generating the Summary Graph}
	\label{alg:summarygraph}
	\small
	{\scalefont{0.85}
		\SetKwFunction{succ}{Succ}
		\SetKwFunction{entry}{Entry}
		\SetKwFunction{exit}{Exit}
		\SetKwFunction{getNode}{GetNode}
		\SetKwFunction{entriesCall}{entriesCall}
		\SetKwFunction{raiseQuery}{RaiseQuery}
		\SetKwInOut{Input}{Input}
		\SetKwInOut{Output}{Output}
		
		\Input{$icfg = \langle N, E \rangle$ ICFG of the API method}
		\Output{$SG = \langle N_s, E_s \rangle$ the summary graph of the API method}
		set $SG$ to $\{\}$\\
		set $worklist$ to $\{\}$\\
		$n_0 = \entry(icfg)$; $q = n_0$ // $n_0$ is also the entry for summary\\ 
		
		add $(n_0,q)$ to worklist
		
		\While{$worklist \neq \{\}$}{
			remove pair $(n, q)$ from $worklist$\\
			\If{$n$ is the marked node or $n$ is the exit of $icfg$}{
				add edge $\left<q, n\right>$ to $SG$ \\
				$newq = n$\\

			}
			\lElse{ $newq = q$}
			
			\lForEach{$s \in \succ{n, icfg}$}{add $(s,newq)$ to worklist}
			
		}
		
		\Return{$SG$}
	}
\end{algorithm}
	\section{Applying PCS}~\label{sec:apply}
In this section, we show how we use generated PCSs to construct inter-callback ICFGs for Android apps and also to detect infeasible callback sequences.


As we mentioned in the introduction, there are two major factors that determine the control flow of the app, GUI and Android API methods. Here, our focus is to sequence callbacks related to the Android API calls in the app. The functionalities of these callbacks include, but are not limited to, the Android lifecycles and component interactions.

To construct the app's inter-callback ICFG, we start at some top level method. It can be, for example, a handler for a GUI event.  We first build the ICFG for this callback. We then traverse the ICFG, and when an Android API call is encountered, we add edges to connect the call site of the API call to the entry of the summary, and also from the exit of the summary back to the call site. Next, we identify the implementation of the callbacks listed in the summary. To do so, we perform a pointer analysis on the app to identify the possible types of the calling object and the actual parameters of the API call. In case the parameter is an \texttt{Intent}, we resolve it to a set of possible component types defined in the app. Based on the types, we find the implementation of the callbacks invoked in the API methods at the call site.

To apply program analysis on the inter-callback ICFGs, we can use predicate nodes and update nodes in the summaries to prune the infeasible paths related to callback sequences. Detecting such infeasible paths can be useful to reduce the number of false positives in static analysis and help better estimate the coverage of testing. 

To compute the infeasible paths on the inter-callback ICFGs of the apps, we implemented a demand-driven branch correlation algorithm~\cite{bodik1997refining}. We raise a query at each predicate node in the summaries. Then, we propagate this query backwards along the paths of the CFG. The query can be resolved at the update node within the same PCS or in a different PCS. The modification of the algorithm here is that the query contains the abstract variables obtained from the predicate nodes, and we need to use the information from the update nodes to resolve the queries.

	\section{Experimental Results}~\label{sec:results}
The goals of our experiments are to show that~1)~the PCSs generated are compact enough to be efficiently used by developers as well as static analysis and testing tools; 2) PCSs can be computed with practical precision and scalability; and 3) PCSs are useful for control flow analysis of Android apps.

\subsection{Experimental Setup}~\label{sec:setup}
We implemented Lithium using Soot \cite{vallee1999soot} for summarizing the Android API methods and for computing the apps' inter-callback ICFGs. To summarize the Android API methods, we used as input the byte code of the Android framework 5.1 implementation and applied \textit{Spark}~\cite{lhotak2003scaling} to build the call graphs for the API methods. To use \textit{Spark}, we built a dummy main method which contains the calls to each of the Android API methods analyzed. To analyze apps, we used the \textit{.apk} files as input and applied Dexpler~\cite{bartel2012dexpler} to convert them to the Soot Jimple representation. 

As we mentioned, the Android framework has millions lines of code and the call graph generated using \textit{Spark} can be imprecise. To reduce the number of false positive callbacks due to call graph imprecisions, we used two heuristics: constraint the size of call chains and constraint the number of possible callers when generating call chains (explained in Section~\ref{sec:identifycallbacks}). The use of these heuristics can make our results unsound, but reduces the number of false positives when the call graph blows up. For our experiments, we use 16 as the maximum length of the call chains and 5 randomly picked callers when traversing the call graph backwards. For analyzing Android apps, we built the call graph for each callback in the app using Class Hierarchy Analysis (CHA).

To perform the experiments, we first evaluated close to 1000 Android apps from the Google Play Market and F-Droid~\cite{fdroid}. Through analyzing the usage of the Android API methods in these apps, we identified 500 frequently invoked Android APIs and generated summaries for them. During summarization, we found a total of 193 PCSs that have at least one node and 127 PCSs have at least one callback. We selected 14 random Android apps from the F-Droid repository and constructed the apps' inter-callback ICFGs using the generated summaries. We also generated dynamic traces through manual and random testing (\textit{Monkey}~\cite{monkey}) to determine whether the paths in the apps' inter-callback ICFGs can be found in real execution traces. 

All of our experiments were run using a virtual machine (VM) with 4 cores and 40GB of memory. The VM runs on a machine with 16 cores of Quad Core AMD Operton 6204. We use a 64-bit JVM with a maximum heap size of 15GB. We provide the detailed experimental results in the next sections. 


\subsection{Compactness of the Summaries}
For all the 500 Android API methods analyzed, we counted the number of nodes in their ICFGs as well as the number of nodes in their PCSs. By comparing the two, we found that the reduction of PCSs over the ICFGs was on average 99\%, and the maximum and minimum reductions found were 100\% and 78\% respectively.
 
We sorted the size of the ICFGs, PCSs, callback nodes, predication nodes and update nodes for the 500 API methods analyzed, and report the minimum, average and maximum number of nodes for each type in rows {\it min}, {\it avg} and {\it max}, under {\it ICFG}, {\it PCS}, {\it Callback}, {\it P-Node} and {\it U-Node} respectively, shown in Table \ref{tab:compact}.  The results show that the size of the PCSs ranges from 0 to 20772 nodes with an average of 330 nodes.

\begin{table}[!ht]\centering
    \caption{Size of ICFGs versus Size of Summary Graphs}\label{tab:compact}
    \begin{tabular}{|l|l|l|l|l|l|} \hline
        & ICFG & PCS & Callbacks & P-Node & U-Node \\ \hline
        min & 2 & 0 & 0 & 0 & 0  \\ \hline
        avg & 58604 & 330 & 55 & 154 & 120 \\ \hline
        max & 208683 & 20772 & 5820 & 14794 & 3807 \\ \hline
    \end{tabular}
\end{table}

\subsection{Correctness of the Summaries}
In this section, we report our studies on the correctness of the PCSs generated by Lithium. In the first study, we compare the PCSs generated from API methods with ICFGs with 1000 nodes or less against a ground truth generated from manual analysis on the source code of the Android framework. From 310 methods that meet this criteria, 23 PCSs have at least one node. Table \ref{tab:accuracy} reports a comparison on a total number of nodes found in the 23 PCSs and their ground truths. Under \textit{Match}, we report the number of nodes that are in both the ground truths and the PCSs. Under \textit{Miss}, we show the total number of nodes that are in the ground truth but missed by our PCS (false negatives). Under {\it Additional}, we list the number of nodes reported in the PCS but not present in the ground truths (false positives). The data show that the precision of the tool is 97\% (just 4 false positives), and the recall is 85\%.
We found 4 callback nodes that were false positives because their object receivers were created in the framework which implies that the objects cannot be passed from any app (internal objects). For the false negatives, we found the main reason to be our heuristics on restricting the size of call chains and the number of callers at each call site to reduce the number of false positives. We found 7 API methods with false negative nodes. The resulting 280 PCSs were proved to not have any node. PCS without nodes can be used by more conservative application call graph analyses such as \textit{Averroes} \cite{ali2013averroes} to reduce the number of edges introduced by possible callbacks. 

\begin{table}[!h]
	\centering
	\caption{Comparison to Manually Identified Ground Truths}\label{tab:accuracy}
	\begin{tabular}{|l||c|c|c|} \hline
		& Match & Missed & Additional \\ \hline\hline 
		Callback & 36 & 8 & 4 \\ \hline 
		P-Node & 95 & 14 & 0 \\ \hline 
		U-Node & 47  & 10 & 0 \\ \hline 
	\end{tabular}
\end{table}

In our second study, we focused on analyzing callback nodes in PCSs with ICFGs with more than 1000 nodes. We took a sample of 300 callbacks nodes from 34 different PCSs to verify their correctness. The precision for this sample was 61\%. The main reason for false positives we found was the imprecise call graphs obtained from the API methods. When we traverse the ICFG of the API method to obtain the call chains for a callback call site, we can encounter a large ICFG that contains a considerable number of virtual functions. The second source of imprecisions involve the object receivers being internal to the framework (the object is created in the framework and it cannot be an object from the app). For example, the PCS for the method \texttt{android.app.Dialog.show} calls callbacks for internal widgets used in windows for dialogs (these objects are internal to the framework). As we mentioned, as future work we consider to implement an analysis to detect all internal objects which can help to improve the precision of the PCSs.

\subsection{Scalability of Generating the Summaries}
\begin{figure}
    \centering
    \begin{subfigure}[b]{0.23\textwidth}
        \includegraphics[width=\textwidth]{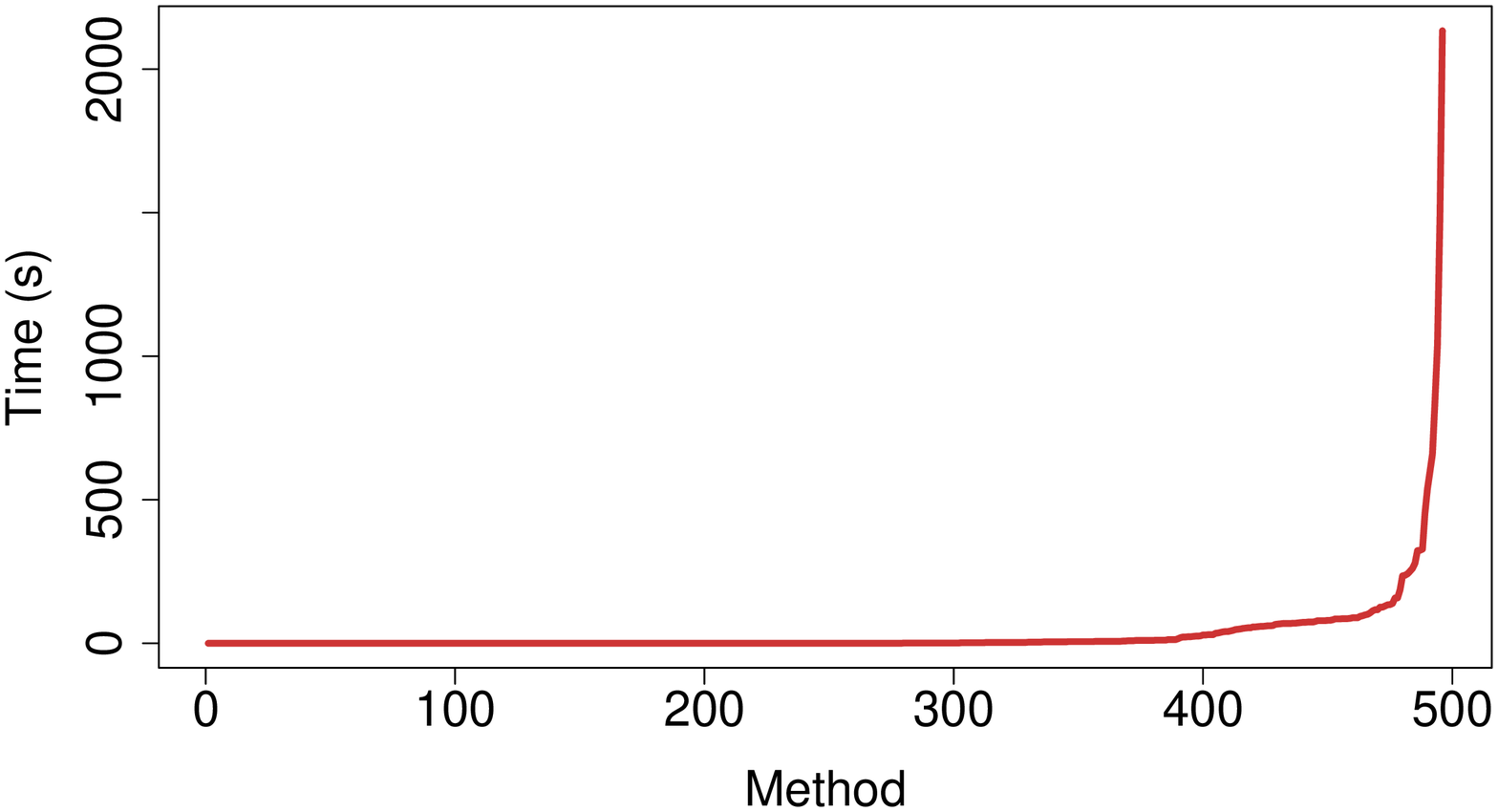}
        \caption{Time in Ascending Order}
        \label{fig:timesummaryasc}
    \end{subfigure}	
    \begin{subfigure}[b]{0.23\textwidth}
        \includegraphics[width=\textwidth]{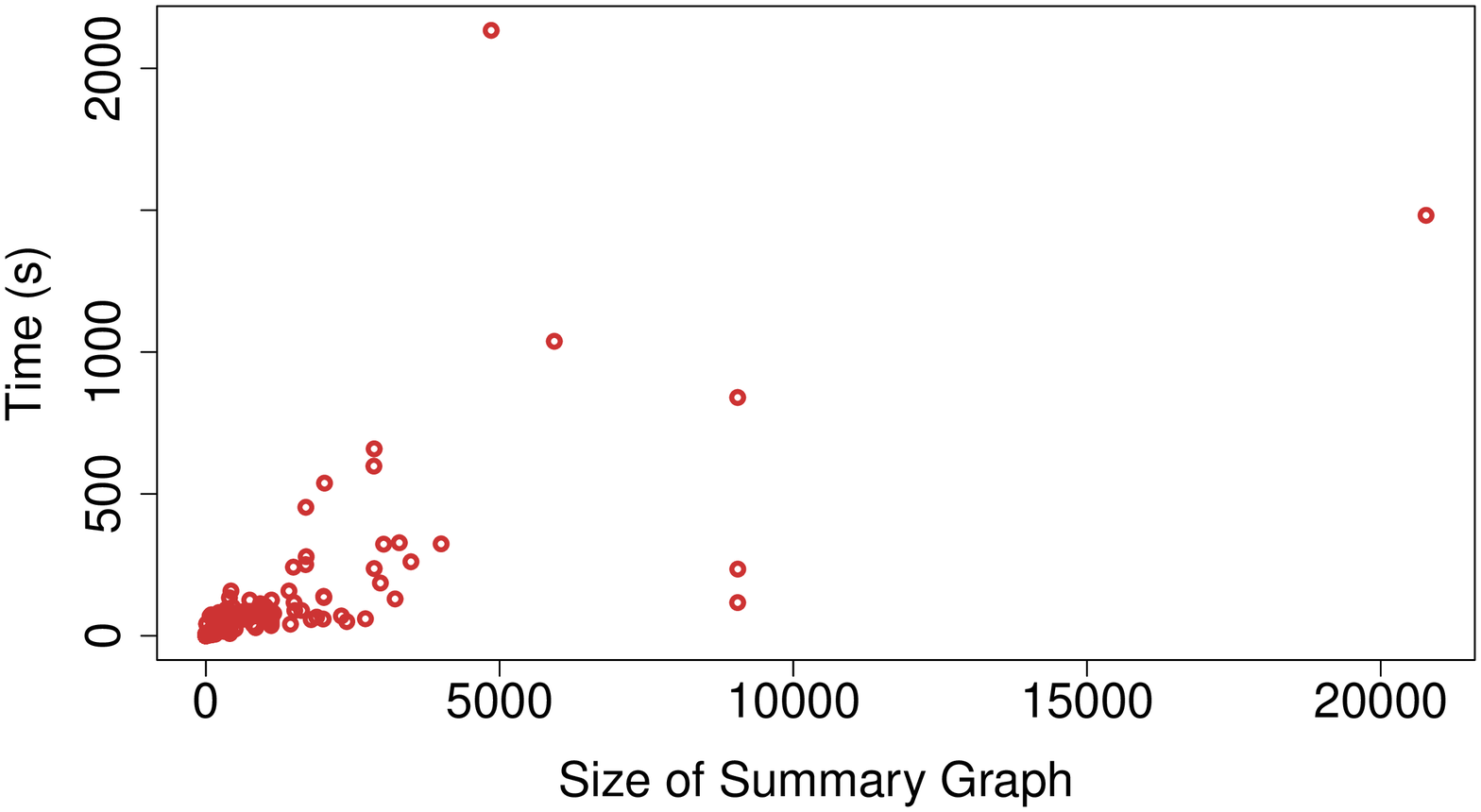}
        \caption{Time vs Size of Summary}
        \label{fig:timevssize}
    \end{subfigure}
    \caption{Scalability of Lithium}
    \label{fig:timesummaries}
\end{figure}

In Figure \ref{fig:timesummaryasc}, we show the time used to build the PCSs for all the 500 API methods in ascending order. For 372 methods, the tool consumed less than 10 seconds having an overall average of 34.5 seconds. One of the aspects that contributes to the scalability of our tool was the demand-driven analysis that identified the call chains of the callbacks and discarded irrelevant methods for the expensive later phases. 

In Figure \ref{fig:timevssize}, we plot the performance against the size of the summary graphs. We inspected the API methods that consumed more than 1000 seconds and found that the backward symbolic substitution analysis (used to resolve abstract variables in predicate nodes and identify update nodes) took more time in these methods because their call chains are longer than the rest of the methods. This behavior can increase the number of paths the analysis has to resolve, therefore, consuming more time. 



\begin{table*}[!ht]\centering
    \caption{Constructing Apps' inter-callback ICFGs Using PCSs} \label{tab:appsanalysis}
    \begin{tabular}{|l|l|l|l|l|l|l|l|} \hline
        \multirow{2}{*}{App} & \multirow{2}{*}{Callbacks} & \multirow{2}{*}{API Calls} & \multicolumn{3}{l|}{inter-callback ICFGs} & \multirow{2}{*}{Longest Path}  & \multirow{2}{*}{Time (s)}	\\ \cline{4-6}
        &  &  & min & ave & max &  &  \\ \hline
        com.blippex.app & 21  & 4 & 1 & 1 & 2 & 2 & 0.2\\ \hline
        net.sourceforge.andsys & 22 & 12 & 1 & 6 & 15 & 9  & 0.1\\ \hline
        com.darknessmap & 25 & 11 & 1 & 2 & 5 & 4 & 0.1\\ \hline
        de.onyxbits.remotekeyboard & 41 & 29 & 1 & 2 & 4 & 5  & 0.2\\ \hline
        com.example.android.contactslist & 44 & 11 & 1 & 8 & 23 & 5 & 0.2\\ \hline
        info.staticfree.SuperGenPass & 50 & 20 & 2 & 4 & 7 & 5 & 0.0\\ \hline
        com.markuspage.android.atimetracker & 65 & 56 & 2 & 4 & 7 & 5 & 0.2\\ \hline
        aarddict.android & 66 & 28 & 1 & 3 & 10 & 13 & 0.3\\ \hline
        de.ub0r.android.websms & 73 & 89 & 0 & 3 & 8 & 6 & 0.3\\ \hline
        com.google.zxing.client.android & 83 & 14 & 1 & 5 & 34  & 84 & 0.2\\ \hline
        a2dp.Vol & 173 & 121 & 1 & 4 & 15 & 15  & 0.6\\ \hline
        org.connectbot & 176 & 108 & 1 & 6 & 27 & 43  & 1.1\\ \hline
         org.openintents.filemanager & 180 & 38 & 1 & 6 & 23 & 44  & 0.6\\ \hline
        com.evancharlton.mileage & 241 & 80 & 1 & 2 & 6 & 5 & 0.4\\ \hline
        
    \end{tabular}
\end{table*}

\subsection{Constructing Apps' inter-callback CFGs with PCSs}
In this section, we report our results on control flow analysis of Android apps using PCSs. In Table \ref{tab:appsanalysis}, under {\it App} we present the name of the 14 apps studied. Under {\it Callback}, we report a total number of callbacks implemented in the apps. Under {\it API calls}, we show the number of Android API calls that were connected to a PCS (using the PCSs generated for 500 API methods). We quantify the number of edges from a callback node in a PCS to the entry point of a callback method under {\it Apps' inter-callback ICFGs}. Since we built partial inter-callback ICFGs for each top level method, we report the average, minimum and maximum number of edges for the partial graphs. We traverse each inter-callback ICFG to find the longest path in terms of callbacks, shown under {\it Longest Path}. Finally, we show the time used (in seconds) to build the inter-callback ICFGs under {\it Time (s)}. We do not include the time Soot took to build ICFGs for each top level method. The construction of the apps' inter-callback CFGs took less than a second for all apps and 0.3 seconds on average. The results show that by modeling API calls using the summaries, we are able to connect callbacks to the control flow graphs with low overhead. The longest callback sequence found was of 44 callbacks for the app {\tt org.openintents.filemanager}. Note that we used the summaries for the 500 most frequently used Android API method, and not all the API calls were modeled. We expect that the larger inter-callback ICFGs and the longer callback sequences would be generated if we plug in more summaries.

\begin{table*}[!ht]\centering
    \caption{Compare Dynamic Traces and Static Paths}\label{tab:traces}
    \begin{tabular}{|l|l|l|l|l|l|l|} \hline
        \multirow{2}{*}{App} & \multicolumn{3}{c|}{Traces} & \multicolumn{3}{c|}{Paths} \\ \cline{2-7}
        & covered-c & total & covered-p & infeasible & in traces & not in traces \\ \hline
        com.blippex & 94\% & 12 & 12 & 0 & 77.8\%  & 22.2\% \\ \hline
        net.sourceforge.andsys & 81\% & 6 & 6 & 0 & 47\% & 53\%\\ \hline
        com.darknessmap & 64\% & 6 & 6  & 0 & 63\% & 37\%\\ \hline
        com.example.android.contactslist & 80\% & 6 & 6 & 12.5\% & 67.5\% & 20\%\\ \hline
        de.onyxbits.remotekeyboard & 66\% & 6 & 6 & 0 & 56.2\% & 43.8\%\\ \hline
        info.staticfree.SuperGenPass & 64\% & 6 & 6 & 20.8\% & 58.4\% & 20.8\%\\ \hline
        com.markuspage.android.atimetracker & 47\% & 6 & 6  & 0 & 41\% & 59\%\\ \hline
        aarddict.android & 65\% & 6 & 6 & 0 & 3\% & 97\%\\ \hline
        de.ub0r.android.websms & 43\% & 6 & 6 & 0 & 21.5 & 78.5\%\\ \hline
        com.google.zxing.client.android & 57\% & 6 & 0  & 0 & 0.3\% & 99.7\%\\ \hline
        org.openintents.filemanager & 35\% & 10 & 10  & 0 & 4.2\% & 95.8\%\%\\ \hline
        org.connectbot & 41\% & 6 & 6  & 0 & 2\% & 98\%\\ \hline
        a2dp.Vol & 47\% & 10 & 9  & 0 & 0.4\% & 99.6\%\\ \hline
        com.evancharlton.mileage & 40\% & 6 & 6 & 0 & 33.8\% & 66.2\%\\ \hline
    \end{tabular}
\end{table*}


In Table \ref{tab:traces}, we show the comparison of the traces generated from testing the apps (see Section \ref{sec:setup}) and the paths (the callback sequences) generated from the inter-callback ICFGs. Column {\it Traces} reports the data obtained from analyzing the traces. Under \textit{covered-c}, we report the percentage of the callbacks covered in the traces. Under {\it total}, we report the total number of runs done for generating traces. Under {\it covered-p}, we report how many of these traces contain the callback sequences obtained from the apps' inter-callback ICFGs. We perform this comparison by verifying whether the paths of callback sequences are subsequences of the dynamic traces. Once we integrate our analysis with a GUI model, we would be able to compare the program paths against the complete traces.   
Our results show that on average, we covered 49\% of the callbacks during testing. We partially covered 96 out of 97 traces from the 14 apps. For app such as {\tt a2dp.Vol}, we found that the tests did not reach some of the API calls we modeled. 
Column {\it Paths} report the analysis of paths. Under {\it infeasible}, {\it in traces} and {\it uncovered}, we report the number of infeasible paths of callbacks detected using the predicate and update nodes; the number of paths of callback sequences that were found in the dynamic traces; and the number of paths of callback sequences that were not found in the traces. 
Among 6 apps, we were able to confirm that more than 50\% of the paths are either infeasible or covered by the traces. Therefore, even with small number of PCSs used in the app analysis, we are able to generate valid paths in some apps. Regarding the apps with low number of paths found in traces, we observed that the traces for these apps had low callback coverage. 



	\section{Discussion}~\label{sec:discuss}
The main source of imprecisions for generating PCSs came from the call graphs generated for API methods. As we mentioned before, they tend to blow up given the complexity of the Android's framework source code. We used the maximum length of call chains and the maximum number of callers as thresholds to restrict the scope of the analysis to reduce the false positives. This sampling heuristics may lead to incomplete numbers of callback, predicate and update nodes. For future work, we plan to use more precise call graph and points-to analysis algorithms to reduce the number of false positives. In addition, our current analysis does not handle callbacks with object receivers passed from different API methods to the framework. We plan to study a solution to resolve such objects during client analysis. 


	\section{Related Work}\label{sec:related}
Our work is closely related to the following two areas. 

\vspace{0.1cm}
\noindent{\bf Inter-callback Analysis of Mobile Apps:}
Many of the analysis tools that need callback sequences of Android apps use manual models to define control flow between the  callbacks \cite{arzt2014flowdroid} \cite{li2015iccta} \cite{yang-icse15} \cite{Hsiao:2014:RDE:2594291.2594330} \cite{6526708} \cite{Yang:2013:AAS:2508859.2516676}. For example, Yang et al. created a graph representation for GUI behaviors in Android apps and used manual constructed models to find which callbacks are invoked when a small subset of API methods are executed (focused on API methods that change the state of GUI elements such as Activities or Dialogs)~\cite{yang-icse15} \cite{yang2015static}. {\it Flowdroid}\cite{arzt2014flowdroid} and \textit{IccTA} \cite{li2015iccta} used the lifecycle of components for inter-callback analysis. The manual models can only handle a subset of the callbacks executed in the Android API methods, and can be imprecise~\cite{wang2016unsoundness} and difficult to update as the Android framework evolves. 
To identify more callback sequences besides manual models, Blackshear \textit{et al.} \cite{blackshear2015selective} introduced the \textit{jumping} framework, which identifies inter-callback control flow constraints via data dependencies between variables located in different callbacks in the apps. Our approach is complementary in that we analyzed the Android framework source code and likely find more callback constraints that are not present in the apps' data dependencies.
Zhang \textit{et al.} \cite{Zhang:2007:ACA:1285240.1285243} proposed an algorithm to resolve library callbacks to improve application call graphs. Their approach uses a data reachability analysis to reduce spurious callback edges for every library call site. This can help to improve the precision when resolving the calling objects of the callbacks; however, they require to analyze the entire library at each call site of the API methods, which can make the analysis intractable even for small apps \cite{Lhotak:2007:CCG:1251535.1251542}. 

\vspace{0.1cm}
\noindent{\bf Pre-computed Summaries:}
EdgeMiner \cite{EdgeMiner} is the closest work to our approach. Their goal is to map Android API methods to callbacks invoked in 
the methods. However, they do not sequence the callbacks. Our PCS technique identified more fine-grained information, including the order of callbacks and the conditions under which the callbacks will be invoked. This helps construct apps' inter-callback ICFGs and also excludes infeasible paths. 
Pre-computed summaries to solve other types of program analysis problems have been defined.
Clapp \textit{et al.} \cite{Clapp:2015:MME:2771783.2771810} mine information flow specifications for the Android API methods. Their specifications identify how
values from the app can be tainted in the Android API methods. Arzt \textit{et al.} \cite{icse16stubdroid} developed a static analysis technique to generate data flow summaries from libraries to use in solving taint analysis problems. However, neither of these two works include any control flow information about callbacks. Rountev \textit{et al.} \cite{Rountev:2006:IDA:2182103.2182107} \cite{rountev2008ide} extended the dataflow summaries generated by interprocedural finite distributive
subset (IFDS) and interprocedural distributive environment (IDE) algorithms to handle callback call sites in the libraries. The summaries are split before and after a callback call site and merged when the target callback method becomes available during client analysis.
Ali \textit{et al.} \cite{ali2012application, ali2013averroes} analyzed the application code using a single summary node to represent all library methods with a less conservative assumption (\textit{separate compilation assumption}).
This approach cannot be directly applied to Android apps as the app depends almost its entire executions on the Android framework, invoking a great number of callbacks and framework method calls. 

	\section{Conclusions and Future Work}~\label{sec:conclusions}
This paper presents a static analysis technique to construct control flow graphs of Android apps related to callback sequences, complementary to the existing GUI models. The novelty of the work is a specification technique named {\it predicate callback summaries (PCS)} designed to model the control flow of callbacks implemented in the API method. We presented the definition, computation and applications of the specification in our framework Lithium. Our experiments reported that using the PCSs generated, we can connect up to 44 callbacks in a path, which previously was only done manually. We are also able to prune infeasible callback sequences, which can improve the precision of static analysis and testing. In the future, we plan to integrate the inter-callback ICFGs we constructed with the control flow graphs constructed by the GUI-based approaches.


    \section{Acknowledgments}
    Danilo Dominguez Perez was supported by IFARHU-SENACYT scholarships from the Government of Panama. We thank Dr. Hridesh Rajan  for the insightful comments.
	
	
	\bibliographystyle{abbrv}
	\bibliography{androidtesting}  

\begin{thebibliography}{10}

\bibitem{handlerthread}
Communicating with the ui thread.
\newblock
  \url{https://developer.android.com/training/multiple-threads/communicate-ui.html}.
\newblock Accessed: 2016-01-25.

\bibitem{fdroid}
F-droid | free and open source android app repository.
\newblock \url{https://f-droid.org/}.
\newblock Accessed: 2015-06-01.

\bibitem{monkey}
Ui/application exerciser monkey.
\newblock \url{https://developer.android.com/studio/test/monkey.html}.
\newblock Accessed: 2016-04-20.

\bibitem{ali2012application}
K.~Ali and O.~Lhot{\'a}k.
\newblock Application-only call graph construction.
\newblock In {\em ECOOP 2012--Object-Oriented Programming}, pages 688--712.
  Springer, 2012.

\bibitem{ali2013averroes}
K.~Ali and O.~Lhot{\'a}k.
\newblock Averroes: Whole-program analysis without the whole program.
\newblock In {\em ECOOP 2013--Object-Oriented Programming}, pages 378--400.
  Springer, 2013.

\bibitem{icse16stubdroid}
S.~Arzt and E.~Bodden.
\newblock {StubDroid}: Automatic inference of precise data-flow summaries for
  the android framework.
\newblock In {\em International Conference for Software Engineering (ICSE)},
  May 2016.

\bibitem{arzt2014flowdroid}
S.~Arzt, S.~Rasthofer, C.~Fritz, E.~Bodden, A.~Bartel, J.~Klein, Y.~Le~Traon,
  D.~Octeau, and P.~McDaniel.
\newblock Flowdroid: Precise context, flow, field, object-sensitive and
  lifecycle-aware taint analysis for android apps.
\newblock In {\em Proceedings of the 35th ACM SIGPLAN Conference on Programming
  Language Design and Implementation}, page~29. ACM, 2014.

\bibitem{bartel2012dexpler}
A.~Bartel, J.~Klein, Y.~Le~Traon, and M.~Monperrus.
\newblock Dexpler: converting android dalvik bytecode to jimple for static
  analysis with soot.
\newblock In {\em Proceedings of the ACM SIGPLAN International Workshop on
  State of the Art in Java Program analysis}, pages 27--38. ACM, 2012.

\bibitem{blackshear2015selective}
S.~Blackshear, B.-Y.~E. Chang, and M.~Sridharan.
\newblock Selective control-flow abstraction via jumping.
\newblock In {\em Proceedings of the 2015 ACM SIGPLAN International Conference
  on Object-Oriented Programming, Systems, Languages, and Applications}, pages
  163--182. ACM, 2015.

\bibitem{bodik1998path}
R.~Bod{\'\i}k and S.~Anik.
\newblock Path-sensitive value-flow analysis.
\newblock In {\em Proceedings of the 25th ACM SIGPLAN-SIGACT symposium on
  Principles of programming languages}, pages 237--251. ACM, 1998.

\bibitem{bodik1997refining}
R.~Bodik, R.~Gupta, and M.~L. Soffa.
\newblock Refining data flow information using infeasible paths.
\newblock In {\em Software Engineering--ESEC/FSE'97}, pages 361--377. Springer,
  1997.

\bibitem{EdgeMiner}
Y.~Cao, Y.~Fratantonio, A.~Bianchi, M.~Egele, C.~Kruegel, G.~Vigna, and
  Y.~Chen.
\newblock Edgeminer: Automatically detecting implicit control flow transitions
  through the android framework.
\newblock In {\em Proceedings of the ISOC Network and Distributed System
  Security Symposium (NDSS)}, 2015.

\bibitem{Clapp:2015:MME:2771783.2771810}
L.~Clapp, S.~Anand, and A.~Aiken.
\newblock Modelgen: Mining explicit information flow specifications from
  concrete executions.
\newblock In {\em Proceedings of the 2015 International Symposium on Software
  Testing and Analysis}, ISSTA 2015, pages 129--140, New York, NY, USA, 2015.
  ACM.

\bibitem{Hsiao:2014:RDE:2594291.2594330}
C.-H. Hsiao, J.~Yu, S.~Narayanasamy, Z.~Kong, C.~L. Pereira, G.~A. Pokam, P.~M.
  Chen, and J.~Flinn.
\newblock Race detection for event-driven mobile applications.
\newblock In {\em Proceedings of the 35th ACM SIGPLAN Conference on Programming
  Language Design and Implementation}, PLDI '14, pages 326--336, New York, NY,
  USA, 2014. ACM.

\bibitem{Lhotak:2007:CCG:1251535.1251542}
O.~Lhot\'{a}k.
\newblock Comparing call graphs.
\newblock In {\em Proceedings of the 7th ACM SIGPLAN-SIGSOFT Workshop on
  Program Analysis for Software Tools and Engineering}, PASTE '07, pages
  37--42, New York, NY, USA, 2007. ACM.

\bibitem{lhotak2003scaling}
O.~Lhot{\'a}k and L.~Hendren.
\newblock Scaling java points-to analysis using spark.
\newblock In {\em Compiler Construction}, pages 153--169. Springer, 2003.

\bibitem{li2015iccta}
L.~Li, A.~Bartel, T.~F. Bissyand{\'e}, J.~Klein, Y.~Le~Traon, S.~Arzt,
  R.~Siegfried, E.~Bodden, D.~Octeau, and P.~Mcdaniel.
\newblock {IccTA: Detecting Inter-Component Privacy Leaks in Android Apps}.
\newblock In {\em Proceedings of the 37th International Conference on Software
  Engineering (ICSE 2015)}, 2015.

\bibitem{dannydig}
Y.~Lin, S.~Okur, and D.~Dig.
\newblock Study and refactoring of android asynchronous programming (t).
\newblock In {\em Automated Software Engineering (ASE), 2015 30th IEEE/ACM
  International Conference on}, pages 224--235. IEEE, 2015.

\bibitem{6526708}
Y.~Liu, C.~Xu, and S.~Cheung.
\newblock Where has my battery gone? finding sensor related energy black holes
  in smartphone applications.
\newblock In {\em Pervasive Computing and Communications (PerCom), 2013 IEEE
  International Conference on}, pages 2--10, March 2013.

\bibitem{pathak2012keeping}
A.~Pathak, A.~Jindal, Y.~C. Hu, and S.~P. Midkiff.
\newblock What is keeping my phone awake?: characterizing and detecting
  no-sleep energy bugs in smartphone apps.
\newblock In {\em Proceedings of the 10th international conference on Mobile
  systems, applications, and services}, pages 267--280. ACM, 2012.

\bibitem{Rountev:2006:IDA:2182103.2182107}
A.~Rountev, S.~Kagan, and T.~Marlowe.
\newblock Interprocedural dataflow analysis in the presence of large libraries.
\newblock In {\em Proceedings of the 15th International Conference on Compiler
  Construction}, CC'06, pages 2--16, Berlin, Heidelberg, 2006. Springer-Verlag.

\bibitem{rountev2008ide}
A.~Rountev, M.~Sharp, and G.~Xu.
\newblock Ide dataflow analysis in the presence of large object-oriented
  libraries.
\newblock In {\em Compiler Construction}, pages 53--68. Springer, 2008.

\bibitem{sridharan2013alias}
M.~Sridharan, S.~Chandra, J.~Dolby, S.~J. Fink, and E.~Yahav.
\newblock Alias analysis for object-oriented programs.
\newblock In {\em Aliasing in Object-Oriented Programming. Types, Analysis and
  Verification}, pages 196--232. Springer, 2013.

\bibitem{vallee1999soot}
R.~Vall{\'e}e-Rai, P.~Co, E.~Gagnon, L.~Hendren, P.~Lam, and V.~Sundaresan.
\newblock Soot-a java bytecode optimization framework.
\newblock In {\em Proceedings of the 1999 conference of the Centre for Advanced
  Studies on Collaborative research}, page~13. IBM Press, 1999.

\bibitem{wang2016unsoundness}
Y.~Wang, H.~Zhang, and A.~Rountev.
\newblock On the unsoundness of static analysis for android guis.
\newblock In {\em Proceedings of the 5th ACM SIGPLAN International Workshop on
  State Of the Art in Program Analysis}, pages 18--23. ACM, 2016.

\bibitem{weiser1981program}
M.~Weiser.
\newblock Program slicing.
\newblock In {\em Proceedings of the 5th international conference on Software
  engineering}, pages 439--449. IEEE Press, 1981.

\bibitem{Weiss:1992:TCC:151333.151337}
M.~Weiss.
\newblock The transitive closure of control dependence: The iterated join.
\newblock {\em ACM Lett. Program. Lang. Syst.}, 1(2):178--190, June 1992.

\bibitem{yang-icse15}
S.~Yang, D.~Yan, H.~Wu, Y.~Wang, and A.~Rountev.
\newblock Static control-flow analysis of user-driven callbacks in android
  applications.
\newblock In {\em International Conference on Software Engineering}, 2015.

\bibitem{yang2015static}
S.~Yang, H.~Zhang, H.~Wu, Y.~Wang, D.~Yan, and A.~Rountev.
\newblock Static window transition graphs for android (t).
\newblock In {\em Automated Software Engineering (ASE), 2015 30th IEEE/ACM
  International Conference on}, pages 658--668. IEEE, 2015.

\bibitem{Yang:2013:AAS:2508859.2516676}
Z.~Yang, M.~Yang, Y.~Zhang, G.~Gu, P.~Ning, and X.~S. Wang.
\newblock Appintent: Analyzing sensitive data transmission in android for
  privacy leakage detection.
\newblock In {\em Proceedings of the 2013 ACM SIGSAC Conference on Computer \&
  Communications Security}, CCS '13, pages 1043--1054, New York, NY, USA, 2013.
  ACM.

\bibitem{Zhang:2007:ACA:1285240.1285243}
W.~Zhang and B.~G. Ryder.
\newblock Automatic construction of accurate application call graph with
  library call abstraction for java: Research articles.
\newblock {\em J. Softw. Maint. Evol.}, 19(4):231--252, July 2007.

\end{thebibliography}
	
\end{document}